\documentstyle[prl,aps,preprint]{revtex}

\newcommand \beq{\begin{eqnarray}}
\newcommand \eeq{\end{eqnarray}}
\def\bfgrad{\mbox{\boldmath$\nabla$}}       
\def\del{\partial}                              
\begin{document}
\draft

\title{Variational calculations with improved energy functionals in gauge theories}
\author{C.~Heinemann$^1$,  E.~Iancu$^2$, C.~Martin$^3$ and D.~ Vautherin$^1$}
\address
      {$^1$LPNHE, LPTPE,  Universit\'es Paris VI/VII, case 127,
F-75252, Paris Cedex 05, France\\
$^2$Theory Division, CERN, CH-1211 Geneva 23, Switzerland\\
$^3$Institut de Physique Nucl\'eaire, F-91406, Orsay Cedex, France}
\date{\today}

\maketitle

\begin{abstract}
For a $SU(N)$ Yang-Mills theory, we present
variational calculations using gaussian wave functionals combined with 
an approximate projection on gauge invariant states.
The projection amounts to correcting the energy of the 
gaussian states by substracting the spurious energy associated
with gauge rotations. Based on this improved energy functional,
we perform variational calculations of the interaction energy
in the presence of external electric and magnetic fields.
We verify that the ultraviolet behaviour of our approximation scheme
is consistent, as it should, with that expected from perturbation 
theory. In particular, we recover in this variational framework
the standard one-loop beta function, with a transparent
interpretation of the screening and anti-screening contributions.
\end{abstract}

]
                            
\narrowtext

\section{Introduction}
\setcounter{equation}{0}

The functional Schr\"odinger picture has proven to be a priviledged tool
in exploring a rich variety of aspects of gauge theories which are beyond the 
scope of perturbation theory \cite{JACKIW}. It is a useful starting
point for developping non perturbative calculations based on the variational
approach. In the case of scalar field theories, static as well as
dynamical variational calculations have been performed by using
trial wave functionals of the gaussian type
\cite{SCHIFF,KUTI,BARNES,BARDEEN,COOPER,PI,EBOLI}.
In the case of gauge theories, some early investigations along
these lines can be found in \cite{GREENSITE,CORNWALL,GOGNY,KERMAN}. However 
the application of variational methods to gauge theories is generally
plagued by the difficulty to implement in a calculable way the requirement
of gauge invariance of physical states \cite{KERMAN,FEYNMAN,KOVNER}. 

Gaussian wave functionals allow for analytic calculations, but are not
gauge invariant except in the Abelian case. In principle, one can 
construct gauge invariant states by averaging gaussian wave 
functionals over all gauge rotations.
This results in an effective non-linear sigma model where 
the fields are the group elements of the gauge transformations \cite{KOVNER}.
However, to make progress with this theory, further approximations
are necessary both in the choice of the kernel of the gaussian, and in
the evaluation of the functional integrals over the gauge group
\cite{KOVNER,BROWN1,BROWN2,ZAREMBO,DIAKONOV}. 
Such approximations, which go beyond the
variational principle, are not always under control.
In particular, in the perturbative regime, they fail to completely reproduce
the one-loop beta function \cite{BROWN1}
(see also \cite{BROWN2,DIAKONOV}). 

In this paper we shall propose a different
 strategy which is inspired by techniques 
developed in 1962 by Thouless and Valatin \cite{THOULESS} to deal with the
restoration of rotational invariance when deformed solutions are obtained in
nuclear Hartree-Fock calculations. 
Rather than using gauge invariant variational states, 
we shall limit ourselves to gaussian wave functionals,
but we shall correct the associated energy functional by a non-local
term, which approximately corresponds to the energy gain
when projecting on gauge invariant states.  In the Abelian case,
this amounts to removing the contribution of the longitudinal
part of the gaussian kernel to the energy. 
In the Yang-Mills theory, the corrective energy term is itself
determined by the variational principle, and the ensuing variational
calculation is a priori non-perturbative.

Our main purpose here is to provide the framework for such a
calculation and verify that, in the perturbative regime (i.e.,
when supplemented with an expansion in powers of $g$), it leads
to results consistent with ordinary perturbation theory. As a 
first check, we shall show that our variational calculation
reproduces the standard one-loop beta function of Yang-Mills
theory, with a transparent interpretation of the
various contributions in terms of screening and anti-screening phenomena.
As a further test, we shall compute the vacuum energy in the presence 
of an external magnetic field $B$, and find that it 
exhibits a minimum at a non vanishing value of $B$, 
in agreement with the perturbative calculation in \cite{SAVVIDY}.

The paper is organized as follows.  In Sec. II we briefly review
the functional Schr\"odinger picture and the variational principle
in field theory.
In Sec. III, we present the Thouless-Valatin formalism and consider,
as a simple illustration,  its application to quantum electrodynamics (QED).
In Sec. IV  we present a variational calculation of the one-loop beta function
which is based on the construction of the interaction energy between external 
electrostatic charges. In Sec. V we consider the energy  of the QCD vacuum
 in the presence of a constant magnetic field.
This provides an alternative computation of the beta function,
and also of the gluon condensate which is found to satisfy
the trace anomaly relation.
A summary of our results and a discussion of 
further possible extensions and applications is presented in Sec. VI.

\section{Variational calculations for gauge fields}
\setcounter{equation}{0}

We consider the functional Schr\"odinger description of
the $SU(N)$ Yang-Mills theory. In the temporal gauge $A^0_a=0$,
the canonical coordinates are the vector 
potentials $A^i_a({\bf x})$ and the electric fields $E^i_a({\bf x})$,
which we shall often write as color matrices in the adjoint
representation: e.g.,  $A^i\equiv A^i_b T^b$
(the color indices $a, \,b,\dots$ run from 1 to $N^2-1$).
The generators $T^a$ of the color group are taken to be
Hermitian and traceless; they satisfy 
\begin{displaymath}
[T^a,T^b]=if^{abc}T^c,~
{\rm Tr}\,(T^aT^b)=N\delta^{ab},~
 (T^a)_{bc}=-if^{abc}.\end{displaymath}
The  Hamiltonian density reads ($g$ denotes the coupling constant) :
\begin{equation} \label{HTG}
{\cal H}({\bf x})\,=\,\frac{1}{2}\left\{g^2E^i_a E^i_a({\bf x})
\,+\,
\frac{1}{g^2}B^i_a B^i_{a}({\bf x})\right\},\end{equation}
with the color magnetic field:
\beq
{\bf B}_a\,=\,{\bfgrad}\times{\bf A}_a\,+\,\frac{1}{2}\,f_{abc}
{\bf A_b \times A_c}.\eeq
Note that our conventions are such that the QCD coupling constant
is absorbed in the normalization of the vector potentials.
With these conventions, the covariant derivative reads 
$D_i\,=\,\del_i - iA_i$, and the electric fields $E^i_a$ are 
canonically conjugate to the vector potentials $A^i_a$:
$[E^i_a({\bf x}), A^j_b({\bf y})] = i\delta^{ij}\delta_{ab}
\delta^{(3)}({\bf x-y})$.

In the Schr\"odinger representation, the states
are represented by functionals of $A^i_a({\bf x})$,
$\Psi[{\bf A}]$, and the electric field is acting
on such states by functional differentiation:
\begin{equation}
 E^i_a({\bf x}) \Psi[{\bf A}]\,=\, i\,\frac{\delta}{\delta A^i_a({\bf x})}
\,\Psi[{\bf A}]\,.\end{equation}
The Hamiltonian $H$ commutes
with the generator ${\cal G}$ of time-independent gauge transformations,
\begin{equation} \label{GL}
{\cal G}({\bf x})\,\equiv\,{\mbox{\boldmath$\nabla$}} \cdot {\bf E}({\bf x})
\,+\,i[A^i,E^i],\end{equation}
so it is possible to diagonalize $H$ and ${\cal G}$ simultaneously.

The physical states are constrained by Gauss' law:
\begin{equation} \label{GAUSS}
{\cal G}({\bf x}) \Psi[{\bf A}]\,=\,0,
\end{equation}
which is the requirement of gauge invariance.
[More precisely, eq.~(\ref{GAUSS}) shows that physical states
must be invariant under ``small'' (i.e.,
topologically trivial \cite{JACKIW}) gauge transformations.
We shall not be concerned with the topological aspects of the
gauge symmetry in what follows.] 
More generally, in the presence of matter fields
represented by an external color source with density
$\rho^a$, Gauss' law gets modified as follows:
\begin{equation} \label{GLRHO}
{\cal G}_a({\bf x}) \Psi[{\bf A}]\,=\,\rho_a({\bf x})\Psi[{\bf A}]\,.
\end{equation}

The ground state of QCD is the eigenstate $\Psi_{vac}$ of $H$ of
minimal energy which satisfies Gauss' constraint (\ref{GAUSS}). 
It can be constructed, at least in principle, by using the Ritz
variational principle, which states that:
\begin{equation} \label{Ritz}
\langle H \rangle \equiv 
\frac{< \Psi | H | \Psi >}{< \Psi | \Psi >}
\ge E_{vac},
\end{equation}
with the minimum achieved for $\Psi =\Psi_{vac}$. Here, $\Psi[{\bf A}]$ 
is any wave functional from the physical Hilbert space (i.e., which
satisfies eq.~(\ref{GAUSS})), and $E_{vac}$ is 
the energy of the ground state $\Psi_{vac}$, 
assumed to be non-degenerate.
In practice, however, one has to restrict oneself
to {\it gaussian} wave functionals, the only ones 
which allow an  analytical computation of $\langle H \rangle$.
These have the form
\begin{equation} \begin{array}{lll}
\label{GTWF}
\Psi_0[{\bf A}]\,&=&{\cal N}^{-1} \,{\rm exp}\left\{-\,\frac{1}{4g^2}
\int {\rm d}^3 x \,{\rm d}^3 y 
\left[A^i_a({\bf x})
- \bar A^i_a({\bf x})\right] \right.
\\
&&
\\
&&
\left.
\qquad\times\left(G^{-1}\right)^{ab}_{ij}
({\bf x, y})\left[A^j_b({\bf y})
- \bar A^j_b({\bf y})\right]\right\},
\end{array}
\end{equation}
where the background field ${\bar A}^i_a({\bf x})$ and the kernel
$G^{-1}$ (with matrix elements $\left(G^{-1}\right)^{ab}_{ij}
({\bf x, y})$) are the variational parameters.
We expect that $\bar A=0$ in the vacuum state, and this is
the case that we shall consider mostly in this paper. Still, non-vanishing
values of $\bar A$ will be also considered, in Sec. V below, in a study
of the vacuum stability in the presence of a background color magnetic
field (in the same spirit as, e.g., in Ref. \cite{SAVVIDY}).

The expectation value of the Hamiltonian density in the gaussian state
$\Psi_0$ is \cite{KERMAN} 
\begin{equation} \label{ENERGY}
\begin{array}{lll}
\langle \Psi_0|{\cal H}({\bf x})| \Psi _0\rangle&=&
\frac{1}{2g^2} {\bar {\bf B}}\cdot {\bar {\bf B}} ({\bf x})
\,+\,\frac{1}{8}{\rm Tr}\,
\langle {\bf x}|G^{-1}|{\bf x}\rangle \\
&&
\\

&+&\,\frac{1}{2} {\rm Tr}\, \langle {\bf x}|K G|{\bf x}\rangle\\
&&
\\
&+&\,\frac{g^2}{8}\,
\left( {\rm Tr}\, \{ S_i T^a \langle {\bf x}|G|{\bf x}\rangle
\} \right)^2
\\
&&
\\
&+&\,\frac {g^2}{4}\,{\rm Tr}\,\left 
\{S^i T^a \langle {\bf x}|G|{\bf x}\rangle
S^i T^a \langle {\bf x}|G|{\bf x} \rangle\right \}.
\end{array}
\end{equation}
In this equation ${\bar {\bf B}}$ is the magnetic field associated to the
center ${\bar {\bf A}}$ and $S^i$ is the spin one matrix whose elements
$(j,k)$ are given by $i \varepsilon_{ijk}$.
The notation Tr in equation (\ref{ENERGY}) implies a summation over 
the discrete (color and spatial) indices. For instance,
${\rm Tr}\,\langle {\bf x}|G^{-1}|{\bf x}\rangle 
\,=\sum_{i,a}\,(G^{-1})^{aa}_{ii}({\bf x},{\bf x})$.
Finally, the operator $K$ is the second derivative
of the classical energy with respect to the center ${\bar A}^a_i$. It 
reads, in matrix notations,
\begin{equation} \label{KDEF}
K= (-i {\bf S} \cdot {\mbox{\boldmath${\cal D}$}})^2 - {\bf S \cdot  \bar B},
\end{equation}
where 
\begin{equation} \label{COV} {\cal D}^i \equiv \partial^i \,-\,i {\bar A}^i
\end{equation}
denotes the covariant derivative defined
by the background field ${\bar A}^i\equiv {\bar A}^i_aT^a$.
In particular, 
\beq
K_{ij}(p)\,=\,p^2\delta_{ij}-p_ip_j\qquad{\rm
for}\,\,\,\,\bar A=0\,.\eeq

In the case of non-Abelian gauge theories, however,  gaussian functionals
like eq.~(\ref{GTWF}) suffer from a major drawback: they do not satisfy the
requirement of gauge invariance (\ref{GAUSS}). It is in principle
possible to construct gauge invariant states by {\it projection},
i.e. by averaging a gaussian functional
over all its gauge transformations.
This is achieved by means of the formula
\begin{equation} \label{PROJECTION}
\Psi[{\bf A}] = \frac{1}{ {\cal N}} \int {\cal D} [U({\bf x})]\,
\Psi_U[{\bf A}]\,,
\end{equation}
where the functional integration is performed over the unitary 
$N\times N$ matrix field $U({\bf x})$, with the adequate group 
invariant measure, 
and ${\cal N}$ is a normalization factor. The integrand in
eq.~(\ref{PROJECTION}) is the gauge-transform of the
gaussian state $\Psi_0[{\bf A}]$:
\begin{equation}\label{GTPSI}
\Psi_U[{\bf A}]= \Psi_0[U{\bf A}U^+ +
i U \mbox{\boldmath $\nabla$ } U^+].
\end{equation}
The expectation value $E_P$ of the energy in the projected state 
is given by the following formula
\begin{equation} \label{EPROJ}
E_P\,=\,\frac{
\int {\cal D} [U({\bf x})] <\Psi_0|H|\Psi_U>}{
\int {\cal D} [U({\bf x})] <\Psi_0|\Psi_U>}\,,
\end{equation}
which should replace eq.~(\ref{ENERGY}) in practical calculations.
Unfortunately,  eq.~(\ref{EPROJ}) cannot be evaluated in closed form
because the functional integral over the group is not gaussian.
Various approximations to eq.~(\ref{EPROJ}) have been 
considered in Ref. \cite{KOVNER,BROWN1,BROWN2,ZAREMBO,DIAKONOV}.
In what follows, we shall propose a different approximation method
which is inspired from techniques used in nuclear physics to calculate
the zero point rotational energy of deformed nuclei 
\cite{THOULESS,VILLARS,PEIERLS1,SCHUCK}. 

\section{Approximate projection}
\setcounter{equation}{0}

The fact that our variational ground state, namely the gaussian
$\Psi_0$ in eq.~(\ref{GTWF}), is not gauge invariant introduces
a spurious degeneracy in the problem: $\Psi_0$ is degenerate with
all its gauge-transforms defined in eq.~(\ref{GTPSI}).
This leads to the existence of spurious excitations of zero energy
which corresponds to rotations of $\Psi_0$ in the
gauge space; these are, of course, the Goldstone bosons associated
to the spontaneous breaking of the gauge symmetry by $\Psi_0$.
Accordingly, the expectation value (\ref{ENERGY})
of the Hamiltonian in the {\it deformed} state $\Psi_0$ includes
unphysical contributions expressing the kinetic energy of the gauge
rotations. The Thouless-Valatin formalism  \cite{THOULESS}
provides us with a method to estimate, and thus subtract away,
such unphysical contributions.

\subsection{The case of rotations}

This formalism is best explained on the example of the collective
rotations of a deformed nucleus. There, the equivalent of our present
variational calculation with gaussian wave functionals is the so-called
{\it Hartree-Fock approximation} where the nuclear wave function is
represented by a Slater determinant formed with $A$ single-particle
wave functions $\varphi_k(x)$ (for $A$ nucleons) 
\cite{SCHUCK}. The latter are determined by solving
the Hartree-Fock equations, i.e., the variational
equations obtained by minimizing the expectation
value of the Hamiltonian in the subspace of Slater determinants.
Although the Hamiltonian $H$ is rotationally invariant,
the Hartree-Fock equations may lead to non-invariant solutions
describing nuclear deformations. If this is the case, then the
spectrum develops a ground state rotational band,
\beq E_J \,=\,E_0 \,+\,\frac{J(J+1)}{2 I}\,,\eeq
where $J(J+1)$ is the eigenvalue of the angular momentum 
operator ${\bf J}^2$ and $I$ is the moment of inertia.
One may then conclude that the operator:
\beq\label{tildeH}
\tilde H\,\equiv\,H\,-\,\frac{{\bf J}^2}{2 I}\eeq
is the Hamiltonian of the {\it intrinsic} (that is, non-rotational)
motion: to the ground state rotational band of $H$ corresponds now
a single eigenvalue $E_0$ of $\tilde H$. Then, eq.~(\ref{tildeH})
provides an approximate separation of the dynamics into intrinsic and
rotational motion, which is reminiscent of the familiar separation 
of the center-of-mass motion by the formula:
\beq\label{tildeHP}
\tilde H\,\equiv\,H\,-\,\frac{{\bf P}^2}{2 M}\,,\eeq
where 
\begin{equation} \label{PTOTAL}
{\bf P}={\bf p}_1 +{\bf p}_2 + \ldots +{\bf p}_A, 
\end{equation}
is the total momentum operator, and $M=Am$ is the nuclear mass. 
Thus, a mean field description of the intrinsic motion can
be given by performing variational (or Hartree-Fock) calculations
for the substracted Hamiltonian $\tilde H$. The only question is,
what is the value of the moment-of-inertia parameter $I$ ?

One can answer this question by studying rotations of the deformed
mass distribution. Assume that the nucleus has axial symmetry 
with respect to the $z$-axis, and consider an uniform rotation with
angular velocity $\omega$ about 
the $x$-axis. If $\Psi_\omega(t)$ is the {\it exact}
(time-dependent) state describing such a rotation, then
\beq\Psi_\omega(t)\,=\,{\rm e}^{-i\omega t J_x}{\rm e}^{-iE_\omega t}
\Psi_\omega(0),\eeq
which satisfies the time-dependent Schr\"odinger equation,
\beq i\frac{\del\Psi_\omega}{\del t}\,=\,H \Psi_\omega,\eeq
provided $\Psi_\omega(0)$ is a solution to the following
time-independent problem:
\beq 
(H-\omega J_x)\Psi_\omega(0)\,=\,E_\omega\Psi_\omega(0).\eeq
This is, of course, just the familiar transformation to the rotating
frame of reference, which leads us to consider
the variational problem for the following Hamiltonian:
\begin{eqnarray} \label{3e1}
H_\omega&=&H-\omega J_x.
\end{eqnarray}
This is equivalent to a {\it constrained} variational problem with
the subsidiary condition that $<J_x>$ has a given value; $\omega$ plays
then the role of the Lagrange multiplier. The constrained Hartree-Fock
calculation determines the optimal independent nucleon wave function
$\Psi_\omega\equiv \Psi_\omega(0)$ in the rotating frame.
(As $\omega \to 0$, $\Psi_\omega \to \Psi_0$ which is
the  deformed Hartree-Fock ground state.) Once the optimal wave function is
found it is possible to obtain an estimate of the moment of inertia $I$ of the
nucleus by considering the limit
\begin{equation} \label{3e2}
I = \lim_{\omega \to 0} \frac{ < \Psi_\omega|J_x| \Psi_\omega>}{\omega}.
\end{equation}
This finally allows us to estimate the zero-point rotational
energy $\Delta E_{TV}$ in the deformed state $\Psi_0$ :
\begin{eqnarray} \label{3e3}
\Delta E_{TV}&=& \frac{< \Psi_0| {\bf J}^2| \Psi_0>}{2I}
\end{eqnarray}
Eq.~(\ref{3e3}) is the expected gain in energy when projecting the deformed
Hartree-Fock ground state $\Psi_0$ onto a rotationally invariant state. 
That is, the corrected average energy after projection, which is the
energy of the intrinsic motion (cf. eq.~(\ref{tildeH})), reads
\begin{eqnarray}\label{EPROJ1}
\tilde E&=&E_0-\Delta E_{TV},
\end{eqnarray}
where $E_0 \equiv <\Psi_0| H|\Psi_0>$.

The Thouless-Valatin method is an approximation which is
expected to be valid for large deformations, or more precisely when the
deformation produces a large expectation value of the square of the angular
momentum in the Hartree-Fock ground state \cite{VILLARS}. Indeed in this case
it can be shown that the projection onto invariant states can be accurately
performed because the overlap between two states differing in their
orientation by an angle $\theta$ is sharply peaked near $\theta$=0 thus
allowing for an an expansion in the vicinity of this point
\cite{PEIERLS1}. 

\subsection{Application to QED}

When going to gauge theories in the variational method, 
the Hartree-Fock ground state $\Psi_0$ is replaced
by the gaussian trial wave functional (cf.  eq.~(\ref{GTWF})),
and the angular momentum operator is replaced by the generator
of the gauge transformations, ${\cal G}^a({\bf x})$ (cf. eq.~(\ref{GL})).

As a first illustration let us consider a variational calculation 
in QED, with the gaussian variational Ansatz
\begin{equation} \label{3e4}
\Psi_0[{\bf A}]= {\cal N}^{-1} \exp \{- <A| \frac{1}{4G}|A> \},
\end{equation}
where ${\cal N}=({\rm det} G)^{1/4}$ and 
the expression in the exponent is a condensed notation for the
convolution
\begin{eqnarray}
<A| \frac{1}{4G}|A> \equiv
\frac{1}{4}
\int {\rm d}^3 x {\rm d}^3 y\,A^i({\bf x})\,G^{-1}_{ij}
({\bf x, y})\,A^j({\bf y}).\end{eqnarray}
This wave functional is gauge invariant provided its kernel $G^{-1}$
is transverse: $\partial_i G^{-1}_{ij}=0$. Let us assume, however,
that this is not the case, and see what the
Thouless-Valatin correction would predict in this case. The operator playing
the role of the angular momentum is the charge density operator
\begin{equation} \label{3e5}
{\cal G}({\bf x})={{\bfgrad}\cdot} {\bf E}({\bf x}),
\end{equation}
and the generalization of the Thouless-Valatin formula (\ref{3e3})
for the energy correction reads
\begin{equation} \label{3e13}
\Delta E_{TV}= \int {\rm d}^3 x {\rm d}^3 y\,
< \Psi_0|{\cal G}({\bf x}) {\cal G} ({\bf y})|\Psi_0 > 
\,< {\bf x} | \frac{1}{2{\cal I}} | {\bf y} >,
\end{equation}
where the "moment of inertia" ${\cal I}$ is now
a matrix in coordinate space, with matrix elements
$<{\bf x}|{\cal I}|{\bf y}>\equiv {\cal I}({\bf x, y})$.
This is obtained 
via a constrained variational calculation with Hamiltonian
$H_\omega=H-H_{ext}$, where
\begin{equation} \label{3e6}
H= \frac{1}{2} \int {\rm d}^3 x \,\{{\bf E}^2({\bf x}) + 
({\bfgrad}\times {\bf A})^2 \},
\end{equation}
and the external constraint
\begin{equation} \label{3e7}
H_{ext} = \int {\rm d}^3 x \,\omega({\bf x}) {{\bfgrad}\cdot} {\bf E} ({\bf x}).
\end{equation}
In the present context, the Lagrange multiplier $\omega({\bf x})$ 
plays the role of the temporal component $A^0({\bf x})$ of the 
gauge vector potential. The solution to this constrained 
variational problem is of the form\footnote{Eq.~(\ref{3e8}) can 
be simply understood by recalling that,
in the presence of a constraint of the form $H_{ext}=\alpha p$,
the ground state wavefunction of an harmonic 
oscillator is modified by a factor $\exp{(ip_0x)}$,
but its width remains unchanged.} 
\begin{equation} \label{3e8}
\Psi_{\omega}[ {\bf A}] = \exp \{- i<{\bf F}|{\bf A}> \} \Psi_0[ {\bf A}],
\end{equation}
where the vector field ${\bf F}({\bf x})$ is a new variational parameter,
which expresses the expectation value of the electric field in the
state (\ref{3e8}), $F^i=<\Psi_{\omega} | E^i| \Psi_{\omega}>$, and
is determined by minimizing
\begin{eqnarray} \label{3e9}
E_{\omega}&\equiv &<\Psi_{\omega} | H - H_{ext} | \Psi_{\omega}>\\
&= &
E_0 + \frac{1}{2} \int  {\rm d}^3x\,({\bf F}+{\bfgrad} \omega)^2 \,-\,
\frac{1}{2} \int  {\rm d}^3x\,(\bfgrad  \omega)^2.\nonumber
\end{eqnarray}
We have denoted here ($V$ is the total volume of the space)
\begin{eqnarray}\label{E0}
E_0\,&\equiv&\,<\Psi_0| H|\Psi_0>\\
&=&\,
\frac{V}{2}\int \frac{{\rm d}^3 p}{(2 \pi)^3}~ \left\{
\frac{1}{4}\, G^{-1}_{ii} ({\bf p})\,+\,(p^2\delta_{ij}-p_ip_j)
G_{ij}({\bf p})\right\}.\nonumber\end{eqnarray}
Note that the magnetic piece of the energy (\ref{E0})
(the second term between  parentheses) involves  only
the transverse components of $G_{ij}$, while the electric piece
involves also its longitudinal component.

The functional $E_{\omega}[{\bf F}]$ in eq.~(\ref{3e9}) attains
its minimum for ${\bf F}= - {\bfgrad}  \omega$, in which case
\begin{equation} \label{3e10}
<\Psi_{\omega} |{{\bfgrad}\cdot}{\bf E} | \Psi_{\omega}>\,=
{{\bfgrad}\cdot} {\bf F}= - \Delta \omega.
\end{equation}
According to (\ref{3e2}), the moment of inertia is obtained as
(with $< {{\bfgrad}\cdot}{\bf E}>_\omega \,\equiv\,
<\Psi_{\omega} |{{\bfgrad}\cdot}{\bf E} | \Psi_{\omega}>$) :
\begin{eqnarray} \label{3e11}
<{\bf x}|{\cal I}|{\bf y}>&= 
&\frac{ \delta < {{\bfgrad}\cdot}{\bf E}({\bf x})>_\omega}
{\delta \omega({\bf y})}\nonumber\\
&=& <{\bf x}|- \Delta |{\bf y}>\,,
\end{eqnarray}
whose inverse is simply the Coulomb propagator:
\begin{equation} \label{3e12}
<{\bf x}| \,\frac{1}{I}\, |{\bf y}>\,=\,\frac{1}{4 \pi | {\bf x}- {\bf y}|}\,.
\end{equation}
We thus obtain the following expression for the
Thouless-Valatin correction (\ref{3e13}) in QED:
\beq \label{3e14}
\Delta E_{TV}&=&\frac{1}{2} \int  {\rm d}^3 x {\rm d}^3 y
\nonumber\\&{}&\,\,<\Psi_0| {{\bfgrad}\cdot}{\bf E}({\bf x})~ 
{{\bfgrad}\cdot}{\bf E}({\bf y})|\Psi_0>
\frac{1}{4 \pi | {\bf x}- {\bf y}|}\,,
\eeq
which is recognized as the electrostatic energy in
the state $\Psi_0$. For the gaussian state (\ref{3e4}), this gives
\begin{equation} \label{3e15}
\Delta {E}_{TV} \,=\, \frac{V}{8}\int \frac{ {\rm d}^3 p}{(2 \pi)^3}~ 
 \frac{p_i p_j}{p^2}\, G^{-1}_{ij} ({\bf p})\,,
\end{equation}
which simply subtracts the longitudinal piece of the electric energy
in eq.~(\ref{E0}).

The corrected energy $\tilde E\equiv E_0 - \Delta E_{TV}$ 
reads therefore (cf. eqs.~(\ref{E0}) and (\ref{3e15}))
\begin{eqnarray}\label{projE}
\tilde E\,=\,V\int \frac{{\rm d}^3 p}{(2 \pi)^3}~ \left\{
\frac{1}{4}\, G^{-1}_T ({p})\,+\,p^2 G_T({p})\right\},\end{eqnarray}
and involves only the transverse piece
$G_T\equiv \frac{1}{2}(\delta_{ij}- \hat p_i \hat p_j)G_{ij}$
of the kernel $G_{ij}$ (we have written here $p\equiv |{\bf p}|$
and $\hat p_i\equiv p_i/p$).
Then, the variational equation $\delta {\tilde E}/\delta{G_T} \,=\,0$
gives $G_T$ in the expected form
\beq\label{GT}
G_T(p)\,=\,\frac{1}{2p}\,.\eeq
Together, eqs.~(\ref{projE}) and (\ref{GT}) yield an energy density
\begin{eqnarray}
\tilde E\,=\,V\int \frac{{\rm d}^3p}{(2 \pi)^3}~|{\bf p}|\,,\end{eqnarray}
which is indeed the correct result for the QED ground state \cite{JACKIW}.
Thus, in the case of QED, the approximate projection method
of Thouless and Valatin correctly subtracts the contribution
of the unphysical, gauge, degrees of freedom from the average energy.
Actually, since QED without fermions is a free theory, the
variational solution above coincides with the exact solution
\cite{JACKIW}: the exact ground state is a gaussian 
wave functional like eq.~(\ref{3e4}) with a transverse kernel
determined by eq.~(\ref{GT}):
\beq\label{QEDGS}
\Psi_{vac}
[{\bf A}]&=&\exp \left\{-
\int \frac{{\rm d}^3 p}{(2 \pi)^3}\,\frac{p}{2}\,
A^i({\bf p})(\delta_{ij}-\hat p_i\hat p_j)A^i(-{\bf p})\right\}
\nonumber\\
&=&\exp \left\{-\int {\rm d}^3 x \,{\rm d}^3 y\,
\frac{{\bf B}({\bf x})\cdot {\bf B}({\bf y})}
{4\pi^2|{\bf x-y}|^2}\right\}.\eeq
Moreover, wave functionals of the type shown in eq.~(\ref{3e8})
--- i.e., gaussian states
with a transverse kernel and a non-trivial phase factor ---
correspond to physical {\it charged} states, i.e., states of the
quantum Maxwell theory in the presence of static, classical, external
sources. Indeed, any such a state (which we denote here as $\Psi_{c}$)
satisfies:
\beq
{{\bfgrad}\cdot} {\bf E}({\bf x})
\Psi_{c}[ {\bf A}] \,=\,\rho({\bf x})
\Psi_{c}[ {\bf A}],\eeq
with the charge density $\rho({\bf x}) = {{\bfgrad}\cdot} 
{\bf F}({\bf x})$. The corresponding energy includes
the Coulomb energy, as expected:
\beq\label{QEDC}
<\Psi_{c}|H|\Psi_{c}>&=&E_0
+ \frac{1}{2} \int  {\rm d}^3x\,{\bf F}^2({\bf x})\nonumber\\
&=&E_0 + \frac{1}{2} \int  {\rm d}^3x \,{\rm d}^3 y\,
\frac{\rho({\bf x})\rho({\bf y})}{4 \pi | {\bf x}- {\bf y}|}\,.
\eeq
Non-Abelian charged states will be considered in Sec. IV.B below.

For other applications and a more complete 
study of the Thouless-Valatin method in  the context of quantum 
field theory, see \cite{HEINEMANN}.

\subsection{Approximate projection in QCD}

Let us now consider the case of non-Abelian gauge theories. The corresponding
"moment of inertia" is now a color matrix defined as the polarization tensor
(cf. eq.~(\ref{GL}))
\begin{equation} \label{INERTIA}
{\cal I}^{ab}({\bf x}, {\bf y}) = 
\frac{\delta < {\cal G}^a({\bf x})>_\omega}{\delta \omega^b({\bf y})} 
\vert_{\omega=0},
\end{equation}
in the presence of an external constraint
\begin{eqnarray} \label{HEXTNAB}
H_{\omega}&=& H-H_{ext}\nonumber\\
H_{ext}&= &\int {\rm d}^3 x~\omega^a({\bf x})  {\cal G}^a({\bf x}).
\end{eqnarray}
The analog of eq.~(\ref{3e3}), i.e. the gain in energy when projecting
a wave functional $\Psi_0[{\bf A}]$ onto
the subspace of gauge invariant states, reads:
\begin{equation} \label{DETV}
\Delta E_{TV}= \int {\rm d}^3 x {\rm d}^3 y
\langle \Psi_0 | {\cal G}^a({\bf x}){\cal G}^b({\bf y})| \Psi_0 \rangle
\langle a,\,{\bf x}|\frac{1}{2 {\cal I}}|b,\,{\bf y}\rangle 
\end{equation}
The energy functional to be used in variational calculations
is therefore 
\beq\label{NLE}
\tilde E=E_0-\Delta E_{TV}.\eeq

It is also possible to perform a projection on a subspace 
with a given color
charge distribution $<{\cal G}^a({\bf x})>_c\,\neq 0$
(The subscript $c$ refers to expectation values over
charged states). In such a case, however, the gauge generator
appearing in the previous correction formula has to be replaced 
by its deviation ${\hat {\cal G}}$ away from the desired value:
\begin{equation} \label{HATG}
{\hat {\cal G}}^a({\bf x})= {\cal G}^a({\bf x})-<{\cal G}^a({\bf x})>_c.
\end{equation}
This modification guarantees that there is no correction for a state 
which is an exact eigenstate of the charge operator.
The functional to be minimized in the subspace of gaussian functionals is thus
\begin{equation} \label{MODIF-E}
{\tilde E}= E_c -E_{\rm ext}-\Delta{\hat E}_{TV},
\end{equation}
where $E= \langle H \rangle_c$, $E_{\rm ext}=\langle H_{\rm ext} \rangle_c$
is the energy of the
external constraint generating the charged state\footnote{That is, an
exact charged state $\Psi_c$ is defined as an eigenstate of $H-H_{\rm ext}$.}, and
\begin{equation} \label{MODIF-ETV}
\Delta{\hat E}_{TV} = \int {\rm d}^3x  {\rm d}^3y
<{\hat {\cal G}}^a({\bf x}){\hat {\cal G}}^b({\bf y})>_c
\langle a, {\bf x}|\frac{1}{2 {\cal I}}|b, {\bf y}\rangle .
\end{equation}
This procedure is again reminiscent of the elimination of the center-of-mass
motion in the mean field description of a composite system of $A$ particles
\cite{THOULESS}. For a system
characterized by a set of single-particle wave functions $\varphi_1$,
$\varphi_2$, ..., $\varphi_A$
the optimal state in the {\it center-of-mass} frame
is obtained by minimizing the functional
\begin{equation} \label{ECM}
{\tilde E}= \langle H \rangle - \frac{\langle {\bf P}^2 \rangle}{2M},
\end{equation}
The Thouless-Valatin prescription for the total mass  
$M$ in eq.~(\ref{ECM}) is to use the
relation $\langle {\bf P} \rangle\equiv M {\bf v}$, where
$\langle {\bf P} \rangle$ is the expectation value
of the total momentum (\ref{PTOTAL}) in the presence of the
external constraint $H_{ext}={\bf v}\cdot{\bf P}$.
This prescription gives the desired result $M=m A$ where $m$ is the mass of
the individual constituents.
In a {\it moving} frame with velocity ${\bf v}$ 
the single particle wave functions become 
\begin{equation} \label{MOVING-ORBITS}
\varphi_i({\bf x}) \to e^{i \chi({\bf x})} \varphi_i({\bf x}),
\end{equation}
with $\chi({\bf x})= m {\bf v}\cdot{\bf x}$. Individual momentum operators
in the moving frame are obtained by the gauge transformation
\begin{equation} \label{MOMENTUM}
\nabla \to \nabla + i (\nabla \chi),
\end{equation}
i.e. ${\bf p}_i$ $\to$ ${\bf p}_i- \langle {\bf p}_i \rangle$. The functional
providing the adequate state at the minimum is 
\begin{equation} \label{MOVING-FUNCTIONAL}
{\tilde E}= \langle H \rangle - {\bf v}\cdot{\bf P}-
\frac{\langle ({\bf P}- \langle {\bf P} \rangle) ^2 \rangle}{2M},
\end{equation}
in agreement with eq.~(\ref{MODIF-E}).
This procedure to implement Gauss's law will be important in Sec. IV when
applied to the calculation of the interaction energy of color charges.

Still in the case of charged states, the chromostatic 
energy $E_{chromo}$ is given, in our approximation scheme, by the
classical chromostatic energy corrected
by the Thouless-Valatin term :
\begin{equation} \label{EC-ETV}
E_{chromo}=\int {\rm d}^3x {\rm d}^3y
\langle {\cal G}^a({\bf x})\rangle_c\langle{\cal G}^b({\bf y})\rangle_c
\langle a, {\bf x}|\frac{1}{2 {\cal I}}|b, {\bf y}\rangle
+\Delta{\hat E}_{TV}
\end{equation}
so that
\begin{equation} \label{EC-ETVb}
E_{chromo}=
\int {\rm d}^3x \, {\rm d}^3y\,\,
\langle {\cal G}^a({\bf x}){\cal G}^b({\bf y})\rangle_c\,
\langle a, {\bf x}|\frac{1}{2 {\cal I}}|b, {\bf y}\rangle\, .
\end{equation}
(This is the analog of using $E_{rot} = <{\bf J}^2>/2I$ as an approximation
for the rotational energy of a deformed nucleus.)
In the case of QED this identification is obvious on equations
like (\ref{3e14}) or (\ref{QEDC}).

To conclude, the central result of this section is the non-local
energy functional (\ref{NLE}) (or (\ref{MODIF-E}) in the case of charged
states) which approximately corrects for the lack of gauge symmetry
when working with gaussian states. This energy functional is the starting
point of the variational method we propose for gauge theories.
Note that the corrective energy term
$\Delta E_{TV}$ in eq.~(\ref{DETV}) is 
a priori of a non-perturbative nature.
Our aim in what follows
is to check the ultraviolet behaviour of this approximation scheme.
We shall thus consider the variational calculations
in the perturbative regime $g\ll 1$.

\section{One-loop beta function from variational calculations}
\setcounter{equation}{0}

In this section, we shall use eq.~(\ref{EC-ETV}) to estimate the
electrostatic energy $E_{chromo}$ of a non-Abelian charged state, up to order
$g^2$ in perturbation theory. This will allow us to recover the
standard expression for the QCD beta function in the one-loop
approximation.

\subsection{Moment of inertia for color rotations}

The first step is the calculation of the moment of inertia for
color rotations, ${\cal I}_{ab}({\bf x,y})$, which enters eq.~(\ref{EC-ETV}).
Unlike QED, where this quantity has been computed exactly
(cf. eq.~(\ref{3e12})), in QCD we shall give only a perturbative estimate
of $I$, valid to the order of interest (i.e., up to order $g^2$).
To this aim, it is sufficient to perform variational calculations
in the vicinity of the perturbative vacuum.

To zeroth order, the vacuum of the Yang-Mills theory is the same
as for the Maxwell theory, namely (cf. eq.~(\ref{QEDGS})):
\begin{equation} \label{GQCD}
\Psi_0[{\bf A}]= {\cal N}^{-1} \exp \left
\{-\left\langle A\left |
 \frac{G^{-1}}{4g^2}\right |A \right\rangle\right \},
\end{equation}
where
\begin{eqnarray} \label{PERTVAC}
(G^{-1})_{ij}^{ab}({\bf k})&=&  \delta_{ab} 
\left( \delta_{ij} - \frac{k_i k_j}{k^2} \right) G^{-1}_{\bf k},
\end{eqnarray}
and $G^{-1}_{\bf k}= 2k$. In the calculation of $I$ below,
we shall never need to go beyond this leading order approximation
for $G_{\bf k}$.

Note that, even with such a transverse kernel,
the functional (\ref{GQCD}) is still not invariant under {\it non-Abelian}
gauge tranformations; that is, this is a deformed state, according to
the terminology in Sec. III. In order to compute its moment
of inertia under color rotations, one has to study the response of
this state to an external constraint of the form (\ref{HEXTNAB}).
The trial wave functional in the presence of this constraint
reads :
\beq \label{PSIOM}
\Psi_{\omega}[{\bf A}]&=& {\cal N}^{-1}\,
{\rm e}^{-i<{\bf F|A}>}
\,\exp \left\{- \left\langle A\left | \frac{G^{-1}}{4g^2}-i\Sigma
\right |A \right\rangle \right\},\nonumber\\&{}&
\eeq
which involves two additional variational parameters: the vector 
field $F^i_a({\bf x})$
(which fixes the expectation value of the electric field:
$\langle{\bf E}_a\rangle = {\bf F}_a$), and the matrix $\Sigma_{ab}^{ij}
({\bf x, y})$, which is taken to be transverse
in its spatial indices. The emergence of $\Sigma$ is
a hallmark of the non-Abelian behaviour (recall that $\Sigma=0$ in the
corresponding Abelian problem; cf. eq.~(\ref{3e8})).

We shall shortly see that, as $\omega \to 0$, $F$ and $\Sigma$
are linear in $\omega$, while the first correction to $G^{-1}$
is only quadratic\footnote{Such quadratic corrections occur since,
in contrast to QED, the non-Abelian constraint (\ref{HEXTNAB})
generates a coupling  between $G$ and $F$ in the variational equations.}, 
and therefore does not matter for the 
calculation of ${\cal I}$ (cf. eq.~(\ref{INERTIA})). 
Thus, for the present purposes, we can take $G^{-1}$ as in
eq.~(\ref{PERTVAC}). The variational parameters in (\ref{PSIOM})
are obtained by minimizing the following functional
\beq\label{EOMEGA}
E_{\omega} &\equiv& <\Psi_{\omega}|H-H_{ext}|\Psi_{\omega}>
\nonumber\\
&=&  <H>_\omega - \small
\int {\rm d}^3 x~\omega^a <{\cal G}^a>_\omega,\eeq
with respect to variations in $F^i_a$ and  $\Sigma_{ab}^{ij}$.
A straightforward calculation yields:
\beq\label{HOMEGA}
<H>_\omega&=&\frac{g^2}{2}\,\int {\rm d}^3x\Bigl\{
 {\bf F}_a \cdot {\bf F}_a({\bf x}) + 4g^2{\rm Tr}\langle {\bf x}|
\Sigma G\Sigma |{\bf x}\rangle\Bigr\},\nonumber\\&{}&
\eeq
where we have kept only the terms involving the variational
parameters. Similarly,
\beq\label{GOMEGA}
<{\cal G}^a({\bf x})>_\omega \,=\,{\bfgrad}\cdot {\bf F}^a({\bf x})
- ig^2 {\rm Tr}\langle {\bf x}| T^a[\Sigma,G]|{\bf x}\rangle.\eeq
After inserting these expressions in eq.~(\ref{EOMEGA}), and
taking variations with respect to $F$ and $\Sigma$, we derive the
following expressions for the variational parameters
(in momentum space) :
\begin{equation} \label{FFF}
F^i_a({\bf q})=-\,\frac{iq^i}{g^2}\,\omega_a({\bf q}),
\end{equation} 
and, for the transverse\footnote{By which we mean transversality
with respect to both ${\bf k}$ and ${\bf k}'$, 
as requested by the expressions (\ref{HOMEGA}) and
(\ref{GOMEGA}).} components of $\Sigma$,
\begin{eqnarray} \label{SIGMA}
\langle k'\ b|\Sigma|k\ c\rangle &=&\frac{(2\pi)^3 }{V}
\delta({\bf k}'-{\bf k}-{\bf q}) \\
&\times&
\frac{1}{2g^2}\,\omega^a({\bf q}) f_{abc} 
\left(\frac{G_{\bf k}-G_{{\bf k}'}}{G_{\bf k}+G_{{\bf k}'}}\right)\nonumber,
\end{eqnarray}
where $V$ is the total volume and $G_{\bf k}$ is defined in (\ref{PERTVAC}).

By using eqs.~(\ref{GOMEGA}), (\ref{FFF}) and (\ref{SIGMA}), we can
finally express the average color charge $<{\cal G}>_\omega $ in terms
of $\omega^a$. This is conveniently
decomposed into an ``Abelian'' and a ``non-Abelian'' piece,
as corresponding to the two pieces in the r.h.s. of eq.~(\ref{GOMEGA}):
$<{\cal G}^a>_\omega = \rho^a_A +\rho^a_{NA}$, with:
\begin{equation} \label{RHOA}
\rho^a_{A}({\bf q})=i {\bf q}\cdot
{\bf F}^a({\bf q})=\,\frac{1}{g^2}\,{\bf q}^2  \omega ^a({\bf q}),
\end{equation}
and, respectively,
\begin{eqnarray} \label{6e2}
\rho_{NA}^a({\bf q})&=& g^2f_{abc} 
\int \frac{{\rm d}^3k}{(2\pi)^3}\,
(\delta_{ij}-\frac{ k'_i k'_j}{ k'^2})(\delta_{ij}-\frac{ k_i k_j}{ k^2})
\nonumber\\&{}&\,\,\,\,\times(G_{{\bf k}}-G_{{\bf k'
}}) < {\bf k'} c|\Sigma |{\bf k} b>,
\end{eqnarray}
where ${\bf k}'={\bf k}+{\bf q}$. From eqs.~(\ref{SIGMA})--(\ref{6e2})
we note that the ``non-Abelian''
charge density in eq.~(\ref{6e2}) is a correction of order $g^2$ 
relative to the ``Abelian'' one in eq.~(\ref{RHOA}). Thus, as
anticipated after eq.~(\ref{PERTVAC}), it is
consistent to evaluate this correction with 
the free kernel $G_{\bf k}=1/2k$.

The integral in eq.~(\ref{6e2}) is logarithmically ultraviolet
divergent, so it must be evaluated with an upper cutoff.
It turns out that this divergence is a part of the charge
renormalization in QCD (see Sec. IV.B below). To reconstruct the
associated beta function, we need, as usual, only the coefficient 
of the divergent logarithm. The latter is insensitive to the
details of the UV regularization, so we shall consider, for simplicity,
a sharp momentum cutoff $\Lambda$.

Also, in order to isolate the
leading logarithm, we can perform kinematical approximations relying
on the inequality $k \gg q$ (since the external momentum $q$
is fixed, while the leading contribution to the integral in
eq.~(\ref{6e2}) comes from relatively large momenta). Physically, 
we are indeed interested in smooth charge distributions.
This allows us to replace ${\bf k}+{\bf q}$ by ${\bf k}$ and thus
$(\delta_{ij}-\frac{ k'_i k'_j}{ k'^2})(\delta_{ij}-\frac{ k_i k_j}{ k^2})$
by 2. By also using eq.~(\ref{SIGMA}) for $\Sigma$, we then obtain : 
\begin{equation} \label{RHOB}
\rho^a_{NA}({\bf q})=\,
\frac{C_N}{2} \,\omega ^a({\bf q}) X({\bf q}),
\end{equation}
where $C_N={\rm Tr} (T^3T^3)=N$ for $SU(N)$,
and
\begin{equation}
X({\bf q})\equiv \int \frac{{\rm d}^3k}{(2\pi)^3}
\frac{[\varepsilon({\bf k}')-\varepsilon({\bf k})]^2}
{\varepsilon({\bf k}) \varepsilon({\bf k}')
(\varepsilon({\bf k}) + \varepsilon({\bf k}'))}\,,
\end{equation}
with $\varepsilon({\bf k})\equiv |{\bf k}|$.
Here again we can replace ${\bf k}' \simeq {\bf k}$ everywhere
except in the numerator which must be expanded to second order
in $q$:
\begin{equation}
[\varepsilon({\bf k}+{\bf q})-\varepsilon({\bf k})]^2 
\simeq q^2 \cos ^2 \theta.
\end{equation}
where $\theta$ is the angle between the space vectors 
${\bf k}$ and ${\bf q}$. The angular average yields
$\langle \cos^2 \theta\rangle=\int_o^\pi d\theta \sin\theta\cos^2 \theta
= 2/3$, so, finally:
\begin{equation}\label{X1}
X({\bf q})\,=\,\frac{q^2}{12 \pi^2}
\int  \frac{{\rm d} k }{k}\,,
\end{equation}
which is logarithmically divergent in the ultraviolet, as expected, but
also in the infrared: the infrared divergence
is an artifact of the previous manipulations (in a more careful 
calculation, this would be screened by $q$), and to the order
of interest we can just regulate it with an ad-hoc infrared cutoff $\mu$.
This yields $\rho^a_{NA}=\alpha \rho^a_{A}$, with
\begin{equation}\label{alpha}
\alpha\,\equiv\,\frac{g^2 C_N}{48 \pi^2} \ln \frac{\Lambda^2}{\mu^2}\,.
\end{equation}
The resulting value of the moment of inertia ${\cal I}_{ab}
\equiv \delta\rho^a/\delta\omega^b$ reads finally:
\begin{equation}\label{IQCD}
{\cal I}_{ab}({\bf q})=\frac{\delta_{ab}}{g^2}\,q^2(1 + \alpha)\,.
\end{equation}
This should be compared to the corresponding Abelian
 result\footnote{The overall factor $1/g^2$ in eq.~(\ref{IQCD}) is
simply a consequence of our different normalizations for the field strengths
in QCD and QED; compare, in this respect, eqs.~(\ref{3e4}) and
(\ref{GQCD}).} $I(q) = q^2$. We see that the 
quantum fluctuations in QCD produce an increase of the moment of inertia,
which corresponds to the {\it screening} of color charges by
quantum fluctuations. The size of the screening effect 
that we have obtained agrees indeed
with the results of other approaches \cite{TDLEE}.


\subsection{Interaction energy in the presence of a background electric field }

Interesting properties of the vacuum include its response to an external
chromo-electric field, which can be
generated by an external constraint of the form
\begin{equation} \label{EXTERNAL-E}
H_{\rm ext}= g^2 
\int {\rm d}^3x\ {\bf E}_{{\rm ext}}^a({\bf x})\cdot{\bf E}^a({\bf x}).
\end{equation}
For this constraint, we shall compute the induced electric mean field
and charge density, and the associated electrostatic energy. By comparing
the latter with the bare Coulomb interaction,
we shall then identify the chromo-electric susceptibility,
or charge renormalization. As we shall see, the variational formalism provides
a transparent picture of the underlying phenomena of screening and
anti-screening.

The optimal state $\Psi_c$ in the presence of
the constraint (\ref{EXTERNAL-E}) is of the form 
\beq \label{ETATCHARGE}
\Psi_c[{\bf A}]&=& {\cal N}^{-1}\,
{\rm e}^{-i<{\bf F|A}>}
\,\exp \left\{- \left\langle A\left | \frac{G^{-1}}{4g^2}
\right |A \right\rangle \right\},\nonumber\\&{}&
\eeq
where the parameter $F^a_i$ (the electric mean field)
will be related to the external field
$E^i_{\rm ext}$ by the variational equations (see eq.~(\ref{FMIN}) below).
Note that, in contrast to eq.~(\ref{PSIOM}), there is no $\Sigma$ term
in eq.~(\ref{ETATCHARGE}) above; this is so because the external perturbation
here is different (compare eqs.~(\ref{EXTERNAL-E}) and (\ref{HEXTNAB})):
it contains a term linear in ${\bf E}^a$, but no non-Abelian term like
$[A^i,E^i]$. This situation is analogous to the case of an anharmonic oscillator with an external 
constraint $H_{ext}=\alpha p$. The only changes are a factor $e^{ip_0 x}$ and, in higher orders in $g$, 
a modification of the real part of the width.  

According to the discussion in Sec. III
(see especially eqs.~(\ref{MODIF-E}) and (\ref{MODIF-ETV})), the energy
functional to be minimized in this case is
$\tilde E
 = E_c-\langle H_{\rm ext} \rangle_c-\Delta{\hat E}_{TV}$.
The terms involving $F^i_a$ in this functional read:
\begin{eqnarray}  \label{DENA}
E_c-E_{\rm ext}-\Delta{\hat E}_{TV}
&=&{g^2} \int {\rm d}^3x\left\{
\frac{1}{2} \,{\bf F}^a \cdot {\bf F}^a
- {\bf E}_{ext}^a \cdot{\bf F}^a\right\}\nonumber
\\ 
&-& g^2 f_{acd} f_{bef} \int {\rm d}^3x {\rm d}^3y\,
 F_i^d({\bf x})   F_j^f({\bf y})\nonumber
\\&{}&\,\,\times
\langle {\bf x}|G_{ij}^{ce}|{\bf y}\rangle  
\langle a, {\bf x}|\frac{1}{2 {\cal I}}| b,{\bf y}\rangle
\end{eqnarray} 
The last term in the r.h.s. corresponds to the Thouless-Valatin
correction, eq.~(\ref{MODIF-ETV})). Note that, because of the substraction
of the average charge in $\hat{\cal G}^a\equiv 
{\cal G}^a-<{\cal G}^a>_c$,
it is only the non-Abelian part of the Gauss operator
(i.e., the term $i[A^i,E^i]$ in eq.~(\ref{GL}))
which contributes to eq.~(\ref{DENA}).
To evaluate this contribution, we first rewrite it in momentum space:
\begin{eqnarray} \label{EHAT-TV}
\Delta{\hat E}_{TV}&=& g^2f_{acd} f_{bef}\int \frac{{\rm d}^3k}{(2 \pi)^3}  
\frac{{\rm d}^3q}{(2 \pi)^3}\,
F_i^d({\bf q}) F_j^f(-{\bf q})\nonumber\\
&{}& \qquad\,\,\,\,\times\,
G_{ij}^{ce}({\bf k})\,
\langle a |\frac{1}{2{\cal I}( {\bf k}+  {\bf q})}|b \rangle.
\end{eqnarray}
To the order of interest, we can replace the moment of inertia
in the equation above by its leading order expression:
${\cal I}({\bf k})\simeq k^2/g^2$. Then, by performing
similar manipulations as in the calculation of the $X$ term
in eqs.~(\ref{6e2})--(\ref{X1}), we finally obtain:
\beq \label{EHAT-P}
\Delta{\hat E}_{TV}&\simeq&\frac{g^4C_N}{3} \int \frac{{\rm d}^3k}{(2 \pi)^3}  
\frac{{\rm d}^3q}{(2 \pi)^3}\,
F_i^a({\bf q}) F_i^a(-{\bf q}) \,\frac{1}{2k}
\frac{1}{k^2}\nonumber\\
&\equiv& \frac{g^2 \delta}{2} \int {\rm d}^3x\,
F_i^a({\bf x}) F_i^a({\bf x}),\eeq
where
\begin{equation} \label{DELTA}
\delta\,\equiv\, \frac{g^2 C_N}{6 \pi^2}
\int \frac{dk}{k}\,=\,
 \frac{g^2 C_N}{12 \pi^2} \ln \left( \frac{\Lambda^2}{ \mu^2} \right).
\end{equation}
After inserting (\ref{EHAT-P}) in (\ref{DENA})
and minimizing with respect to  $F_i^a({\bf x})$, we obtain
\begin{equation} \label{FMIN}
(1-\delta) F^i_a({\bf x})= E_{{\rm ext}\,a}^i({\bf x}).
\end{equation}

At this point it is convenient to introduce the charge distribution associated
to the external field $E_{\rm ext}^i$,
\begin{equation} \label{RHOEXT}
\rho^a_{\rm ext}({\bf x})\equiv \bfgrad\cdot {\bf E}^a_{\rm ext}({\bf x}),
\end{equation}
to be referred to as the {\it external charge} in what follows:
this would be the charge in the system in the absence of
polarization effects. The actual charge is rather
\beq \label{RHO-RHOEXT}
\rho^a({\bf x})&\equiv&
 \langle \Psi_c | {\cal G}^a({\bf x}) | \Psi_c \rangle\,=\, \bfgrad\cdot {\bf F}^a
\nonumber\\&=&
\frac{\rho^a_{\rm ext}}{1-\delta}\,\simeq\,\rho^a_{\rm ext}({1+\delta}),
\eeq
where the second line follows from eq.~(\ref{RHOEXT}).
Note that this relation implies an {\it antiscreening} of the external 
charge, since $\rho^a$ is bigger than $\rho^a_{\rm ext}$.
The difference \\ $\rho - \rho_{\rm ext} = \rho_{\rm ext}\,\delta$
may be interpreted as an {\it induced charge} 
(see also Sec. IV.C in Ref. \cite{VIKI} and Appendix A for an alternative computation
of this quantity).

We are finally in position to compute the chromostatic interaction $E_{chromo}$
in the optimal state $\Psi_c$. This is given by eq.~(\ref{EC-ETV})
which, together with the above expressions (\ref{RHO-RHOEXT})
 for $\langle {\cal G}^a({\bf x}) \rangle_c$,  and (\ref{EHAT-P})
for $\Delta{\hat E}_{TV}$, implies:
\beq\label{ECQCD}
E_{chromo}&\simeq& g^2 \,\frac{1+3\delta}{1+\alpha}\int \frac{{\rm d}^3q}
{(2 \pi)^3} \,
\rho_{\rm ext}^a({\bf q}) \rho_{\rm ext}^a(-{\bf q})\,
 \frac{1}{2 {\bf q}^2}\,,\nonumber\\&{}&\eeq
up to corrections of higher order in $g$.
There is here a factor $(1 + \delta)^2\simeq 1+2\delta$
arising from the induced charge (cf. eq.~(\ref{RHO-RHOEXT})),
another one arising from the Thouless-Valatin correction (\ref{EHAT-P}),
and a factor $(1 +\alpha)$ due to the moment of inertia (cf. eq.~(\ref{IQCD})).
The interaction energy (\ref{ECQCD}) is still Coulomb like,
\begin{equation} \label{COULOMB}
E_{chromo}= g_R^2 \int \frac{{\rm d}^3q}{(2 \pi)^3} \,
\rho^a_{\rm ext}({\bf q}) \rho^a_{\rm ext}(-{\bf q}) \,\frac{1}{2 {\bf q}^2}\,,
\end{equation}
but with a modified coupling constant given by
\begin{equation} \label{GRUN}
g_R^2(\mu) =g^2 \frac{1+3\delta}{1+\alpha}
\end{equation}
or, to first order in $g^2$,
\begin{equation}
\frac{1}{g_R^2(\mu)} = \frac{1}{g^2} \,-\, 
\frac{11 C_N }{48 \pi^2}\,
\ln \frac{\Lambda^2}{\mu^2}\,.
\end{equation}
This is the correct one-loop value for the renormalized coupling constant 
\cite{TDLEE}. Note that, in the present calculation, this involves
three types of contributions: indeed, the factor $11$ in the last equation
has arised as $11=8+4-1$, where the $8$ corresponds to anti-screening
by the induced charge (cf. eq.~(\ref{RHO-RHOEXT})), the 4 is another
anti-screening contribution due to the Thouless-Valatin correction
(\ref{EHAT-P}), and the $(-1)$ is a screening contribution arising
via the correction of order $g^2$ to the moment of inertia.

\section{Vacuum energy in a magnetic background field}
\setcounter{equation}{0}

In the previous section, we have studied the electric sector of the
vacuum of the Yang-Mills theory, by using a combinaison of variational 
and perturbative techniques. In what follows, we shall perform a similar
analysis of the magnetic sector. To this aim, we consider the Yang-Mills 
theory in the presence of a (constant) magnetic background field 
$\bar B^i_a$, and compute the background field energy
by using the variational principle. The final result is not new
(it coincides with the one-loop result by Savvidy
\cite{SAVVIDY}), but it rather serves as a test for our variational
method in the magnetic sector, and in the perturbative regime.

The relevant trial wave variational is the gaussian 
functional $\Psi_0$ in eq.~(\ref{GTWF}) with the ``center'' 
field $\bar A^i_a({\bf x})$ chosen so as to reproduce the
desired magnetic field $\bar B^i_a$ (a convenient choice
will be given later). This state is not gauge-invariant, so its energy 
$E_0\equiv \langle \Psi_0 |H| \Psi _0\rangle$, eq.~(\ref{ENERGY}),
must be corrected with the Thouless-Valatin energy 
$\Delta E_{TV}$, to be computed in the next subsection.
Then, by applying the variational principle to the corrected
energy $\tilde E=E_0-\Delta E_{TV}$, we shall determine the kernel
of the Gaussian (in  Sec. V.B). Finally, in Sec. V.C, we shall compute
the energy of the magnetic field and the associated gluon condensate, and 
verify that these quantities are related by the trace anomaly 
relation, as it should. In this calculation, the standard one-loop
beta function will emerge once again.

\subsection{The Thouless-Valatin energy in the background field}

We start by computing the moment of inertia ${\cal I}_{ab}({\bf x,y})$
in the presence of the background field $\bar A^i_a({\bf x})$.
As already explained, this requires constructing
the variational ground state $\Psi_\omega$ for the constrained
Hamiltonian $H_\omega$ in eq.~(\ref{HEXTNAB}).
This state is of the form (compare to eq.~(\ref{PSIOM}))
\beq
\Psi_\omega&=&{\cal N}^{-1}\, e^{-i\,\langle F|A-\bar A\rangle}\,
\nonumber\\ &{}&\,\,\,\,\times\exp\left\{-\Big\langle
A-\bar A\Big |\frac{G^{-1}}{4}
-i\Sigma\Big |A-\bar A\Big \rangle\right\},
\eeq
where the parameters $F^i_a$ and $\Sigma_{ij}^{ab}$
are related to $\omega^a$ by the variational principle
(recall the discussion in Sec. IV.A). It turns out that
the matrix $\Sigma$ will not play any role
in what follows: indeed, below we shall need the moment
of inertia only to leading order in $g$, while $\Sigma$ counts 
starting with order $g^2$ (cf. Sec. IV.A).
We then write, as in eq.~(\ref{EOMEGA}),
\begin{eqnarray}
E_\omega\,=\,<H>_\omega-\int d^3x\,
\omega^a({\bf x})<{\cal G}^{a}({\bf x})>_\omega,
\end{eqnarray}
with $<H>_\omega$ given by eq.~(\ref{HOMEGA}), and
\beq\label{GG}
<{\cal G}^a({\bf x})>_\omega\,=\,
({\cal D}_iF^i)^a({\bf x})+{\cal O}(g^2),\eeq
where $ {\cal D}^i \equiv \partial^i -i {\bar A}^i$ is the covariant
derivative defined by the background field (cf. eq.~(\ref{COV})),
and the neglected terms, of ${\cal O}(g^2)$, would involve $\Sigma$
(cf. eq.~(\ref{GOMEGA})).
The variation with respect to $F^i_a$ yields then
\begin{equation}
{F}^i_a({\bf x})=-\frac{1}{g}({\cal D}_i\omega)^a({\bf x}),
\end{equation}
which differs from the corresponding expression in eq.~(\ref{FFF})
only by the replacement of the ordinary derivative $\del_i$ by
the covariant one $ {\cal D}_i$. Together with eq.~(\ref{GG}),
this provides the moment of inertia to the order of interest:
\begin{eqnarray}
\label{defI}
{\cal I}^{ab}({\bf x},{\bf y})\, \equiv \,
\frac{\delta\langle{\cal G}^a({\bf x})\rangle_\omega}{\delta\omega^b ({\bf y})}
&=&-\frac{1}{g^2}\,({\cal D}_x^2)^{ab}\,\delta^{(3)}({\bf x-y})
+{\cal O}(1)\nonumber\\
&=&\frac{1}{g^2}\langle a,\,{\bf x}|\Pi^2|b,\,{\bf y}\rangle+{\cal O}(1 ).
\end{eqnarray}
We have introduced here
the {\it kinetic momentum} $\Pi_j\equiv
i{\cal D}_j=i\delta_j+{\bar A}_j$, and $\Pi^2\equiv \Pi_j \Pi_j$.

Within the same accuracy, one has also:
\begin{eqnarray}\label{GGG}
\langle \Psi_0 | {\cal G}^a({\bf x}){\cal G}^b({\bf y})| \Psi_0 \rangle
&\approx&\frac{1}{4g^2}\,{\cal D}_{i,{\bf x}}^{ac}\,{\cal D}^{j,{\bf y}}_{db}\,
({G}^{-1})_{ij}^{cd}({\bf x},{\bf y})\nonumber\\
&=&\frac{1}{4g^2}\,\langle a,\,{\bf x}|\Pi_i\, G^{-1}_{ij} 
\Pi_j|b,\,{\bf y}\rangle\,.\end{eqnarray}
We are now in position to compute the 
Thouless-Valatin energy $\Delta E_{TV}$, cf. eq.~(\ref{DETV}):
by combining eqs.~(\ref{defI}) and (\ref{GGG}), one obtains:
\begin{eqnarray}
\label{DTVB}
\Delta E_{TV}\,\approx\,\frac{1}{8}\int d^3 x\,
\langle a,\,{\bf x}|\left(\Pi_i\frac{1}{\Pi^2}\Pi_j\right)
 {G}^{-1}_{ij}|a,\,{\bf x}\rangle\,,
\end{eqnarray}
up to corrections of order $g^2$.

\subsection{The variational equation for $G$}

The improved energy functional $\tilde E=E_0-\Delta E_{TV}$
reads therefore (cf. eqs.~(\ref{ENERGY}) and (\ref{DTVB}))
\begin{eqnarray}\label{PROJB}
\tilde E
&=&\int d^3{\bf x}\
\biggl \{ \frac{1}{2g^2}{\bar B}_i^a({\bf x}){\bar B}^a_i({\bf x})\nonumber\\
&&\hskip 0.5cm \,+\,
\frac{1}{8}\,
\langle a,\,{\bf x}|\Bigl(\delta_{ij}\,-\,\Pi_i\frac{1}{\Pi^2}\Pi_j\Bigr)
 {G}^{-1}_{ij}|a,\,{\bf x}\rangle\nonumber\\
&&\hskip 0.5cm +\,\frac{1}{2}\,{\rm Tr} \,[K\,{G }({\bf x},{\bf x})]
+\,{\cal O}(g^2)\biggr\}. 
\end{eqnarray}
Note that the last two terms in (\ref{ENERGY}) do not contribute to this order.
As obvious from this equation, the Thouless-Valatin correction makes
the kinetical part of the energy {\it covariantly} transverse.
Since the operator $K_{ij}$ is transverse as well (cf.
eq.~(\ref{KDEF})),
\beq\label{K2}
K_{ij}\,=\,\Pi^2\delta_{ij}\,-\,\Pi_i\Pi_j \,+\,2[\Pi_i,\Pi_j]\,,\eeq
it follows that the projected energy (\ref{PROJB}) involves
only the {\it transverse} components of the kernel $G$.
Thus, without loss of generality, we can restrict ourselves
to a (covariantly) transverse kernel in what follows:
\beq\label{TRGTR}
\Pi^i G^{-1}_{ij}\,=\,0\,=\, G^{-1}_{ij}\Pi^j.\eeq
To formalize this, it
is convenient to introduce transverse and longitudinal 
projectors as follows:
\beq
{\hat P}_{ij}\,\equiv\,\Pi_i\,\frac{1}{\Pi^2}\,\Pi_j,
\qquad\qquad {\hat Q}\,\equiv\,1-{\hat P}\,.\eeq
They satisfy :
\begin{eqnarray}
{\hat P}^2_{ij}&=&\Pi_i\frac{1}{\Pi^2}\Pi_k\Pi^k\frac{1}{\Pi^2}\Pi^j
={\hat P}_{ij},
\nonumber\\
{\hat Q}^2_{ij}&=&(1-{\hat P})^2_{ij}=(1-{\hat P})_{ij}={\hat Q}_{ij}.
\end{eqnarray}
Then, a transverse kernel is one satisfying
$G={\hat Q} G {\hat Q}$ (and similarly for $G^{-1}$). For such
a kernel, the variational principle (i.e., the minimization
of $\tilde E$, eq.~(\ref{PROJB}), with respect to $G$)
produces the following gap equation:
\begin{equation} \label{GK}
\frac{1}{4{G}^2}\,\simeq\,K,
\end{equation} 
which determines $G$ to the order of interest. In particular,
as $\bar A \to 0$, $G$ reduces to the free, or Abelian,
expression in eqs.~(\ref{PERTVAC}) and (\ref{GT}).
Thus, the only non-trivial effects which are taken here into
account are those associated with the background field.

Note that eq.~(\ref{GK}) can only be valid at sufficiently high
energy, or small coupling constant: indeed,  the 
operator $K$ admits negative modes \cite{SAVVIDY,NIELSEN,OLESEN}.
We thus assume that an infrared cutoff has been set -
this does not affect the ultraviolet behaviour of the theory,
which is our main interest here.

\subsection{The energy of the background field}

The previous equations provide the optimal gaussian kernel 
for a given background field $\bar A$ and thus
 the effective potential $V(\bar A)$ which is the 
expectation value of the energy in this state. 
The next step in our variational approach is to find 
the minimum of the effective potential. Constructing $V$ 
for an arbitrary background is however a difficult task. For this reason
we now consider a restricted variational
space defined by the following background field :
\begin{equation} \label{CHOIX}
{\bar A}_x=0\  ;\ {\bar A}_y=x B T^3\ ;\ {\bar A}_z=0. 
\end{equation} 
This  corresponds to a constant magnetic field in 
the $z$-direction and in the third color. 

With this choice of the background field, we are now able
to compute the energy (\ref{PROJB}) in the optimal variational state,
which is the gaussian state (\ref{GTWF}) with a transverse
(in the sense of eq.~(\ref{TRGTR}))
kernel $G^{-1}$ satisfying eq.~(\ref{GK}).
The latter equation shows that, at the minimum,
the following identity holds:
\begin{equation} \label{IDE}
{\rm Tr}\{K\langle {\bf x}|G|{\bf x}\rangle\}\,=\,
\frac{1}{4}\, {\rm Tr}\,
{\langle {\bf x}|G^{-1}|{\bf x}\rangle}.
\end{equation}
That is, magnetic and electric
fluctuations have equal energies in our variational ground states,
which is merely the virial theorem in the present context. Thus, 
\begin{eqnarray}\label{PROJBbis}
\tilde E_{min} &\simeq& \int d^3{\bf x}
\biggl \{ \frac{1}{2g^2}{\bar B}_i^a({\bf x}){\bar B}^a_i({\bf x})
+\frac{1}{4}
\langle a,\,{\bf x}| {G}^{-1}|a,\,{\bf x}\rangle\biggr\}.
\nonumber\\&{}&
\end{eqnarray}
This involves the matrix element $\langle {\bf x}|G^{-1}|{\bf x}\rangle$,
which we shall compute in Appendix A by using the Schwinger
proper-time representation (cf. eq.~(\ref{GK})) :
\begin{equation}\label{REPSCHW}
\langle {\bf x}|G^{-1}|{\bf x}\rangle=
\frac{1}{\sqrt\pi}\int_0^{\infty}\frac{dt}{t^{3/2}}\langle {\bf x}|
(1-e^{-tK})|{\bf x}\rangle
\end{equation}
This integral develops ultraviolet divergences as $t\to 0$,
which we shall regularize by shifting
the lower bounds of the integral from $0$ to $1/\Lambda^2$. 
As in the electric case, we are mainly interested in the ultraviolet
renormalization of the energy (\ref{PROJBbis}); to this aim,
it is sufficient to extract the terms which diverge when
$\Lambda\to \infty$ in eq.~(\ref{REPSCHW}).
This is described in detail in Appendix A, from which we quote 
here the final result:
\begin{eqnarray}\label{RES}
{\rm Tr}\langle {\bf x}|G^{-1}|{\bf x}\rangle &=&
(\ldots )\Lambda^4-\frac{11C_N}{48\pi^2}\,B^2
\ln\frac{\Lambda^2}{B}+{\cal O}(g^2),\nonumber\\&{}&
\end{eqnarray}
(the coefficient in front of $\Lambda^4$ is an uninteresting
field-independent number that will be omitted in what follows). 
Remarkably, there is no divergent term in
$\Lambda^2$ (which, for dimensional reasons,
would be necessarily of the form $\Lambda^2 B$):
this is so because of rotational and gauge symmetries
which require the magnetic field to enter only in the scalar product
${\bf S}\cdot{\bf B}_a T^a= B S_z T^3$, whose trace is however zero
(see Appendix A for more details).

Note also the numerical factor in front of the logarithmic divergence
in eq.~(\ref{RES}): this is the factor leading to the correct
one-loop beta function after renormalization (see below).
The projection on (covariantly) transverse gaussian states has 
been crucial in getting this factor right: without this,
we would have obtained a factor $\frac{7}{2}$ instead of the correct factor
$\frac{11}{3}$  (compare in this respect eqs.~(\ref{SANSPRO})  
and (\ref{UT}) in Appendix A).

Finally note that the field strength $B$ appears
as an infrared cutoff in eq.~(\ref{RES}). This is expected from equations 
(\ref{EQ1}) and (\ref{EQ2})  where 
the proper time variable always appears in the combination $tB$. 
However a complete derivation 
of (\ref{RES}) requires a detailed treatment of unstable modes.

To conclude,
\begin{eqnarray} 
&&\tilde E/V\,\simeq\,
\frac{1}{2g^2}\,B^2\,-\,C_N\frac{11}{48\pi^2}
\frac{B^2}{2}\ln\frac{\Lambda^2}{B},
\end{eqnarray}
showing that the background field energy has no field dependent UV 
divergences other than the logarithmic one which can be absorbed into
the renormalization of the coupling constant. We then write, as usual
($\mu$ is the substraction scale),
\begin{equation} \label{beta1l}
\frac{1}{g_R^2(\mu)}=\frac{1}{g^2}\,-\,C_N\frac{11}{48\pi^2}
\ln\frac{\Lambda^2}{\mu^2}
\end{equation}
which provides the correct one-loop beta function, as anticipated.
The renormalized field energy density reads then:
\begin{equation} \label{EOFBSQ}
{\cal H}(B)= \frac{B^2}{2g_R^2(\mu)}+\frac{B^2}{2} \frac{11}{48 \pi^2}
C_N\ln \frac{B}{\mu^2},
\end{equation}
which coincides with the result obtained by Savvidy in perturbation
theory \cite{SAVVIDY}. An advantage of the present approach, however, is
that it can be improved by using a larger variational space,
which is expected to cure the difficulties associated 
with negative modes \cite{NIELSEN,OLESEN}.

As discussed in \cite{SAVVIDY}, the
 energy density (\ref{EOFBSQ}) exhibits a minimum for a
non-zero value $B=B_{min}$ of the background field, with 
\begin{equation} \label{60}
B_{min}=\frac{\mu^2}{\sqrt e} \exp \left(-\frac{16 \pi^2}{g_R^2(\mu)}
\frac{3}{11 C_N} \right) 
\end{equation}
The value of the energy density at this minimum is : 
\begin{equation} \label{61}
  \langle {\cal H} \rangle_{min}=-\frac{1}{64 \pi^2}
  \frac{11}{3} C_N B_{min}^2 
\end{equation}
which is indeed negative. Our variational vacuum state is therefore
characterized by a magnetic field condensate
(see, however, Refs. \cite{NIELSEN,OLESEN} for potential problems
with such a state).

From the previous results, it is now straightforward to evaluate
the gluon condensate in our variational vacuum:
\beq \label{62}
\langle F^{\mu \nu} F_{\mu \nu} \rangle &\equiv &
2 g^2 \left(\frac{1}{g^2}\langle B_i^a B_i^a  \rangle
- \langle g^2 E_i^a E_i^a  \rangle \right) \nonumber\\
&=&2 B_{min}^2,
\eeq
where the second line follows from the aforementioned
``virial theorem'' (\ref{IDE}).
Eqs.~(\ref{61}) and (\ref{62}) can be combined into:
\begin{equation}\label{66}
\langle {\cal H} \rangle_{min} = -\frac{11 C_N}{128 \pi^2} 
\,\langle F^{\mu \nu} F_{\mu \nu} \rangle_{min} \, \end{equation} 
which is consistent, as it should,
with the trace anomaly relation:
\begin{equation} \label{67} 
\langle \theta^{\mu}_{\mu}\rangle = \frac{\beta(g)}{2 g^3} \: \langle 
F^{\mu \nu} F_{\mu \nu} \rangle. \end{equation} 
Indeed, with $\langle {\cal H} \rangle =
\frac{1}{4} \langle \theta^{\mu}_{\mu}\rangle$, and the one-loop
beta function (which here is a consequence of eq.~(\ref{beta1l}))
\begin{equation} \label{68} 
\beta(g)=-\frac{11}{48 \pi^2} C_N g^3 \ , \end{equation}
eq.~(\ref{67}) becomes identical to eq.~(\ref{66}).

An attractive feature of the formula (\ref{66})  is that it involves two quantities which are independently
accessible experimentally (at least indirectly). 
Indeed, the left hand side of this equation is the energy density of the 
vacuum, which can be identified with the fourth power of the bag constant,
 ${\cal B}^4=-(240MeV)^4$ \cite{CHODOS},
 whereas the right hand side depends on the gluon 
condensate which is known from Ref.
  \cite{SHIFMAN} to be $0.5 GeV^4$. These values are 
compatible with eq.~(\ref{66}) within a 20 percent accuracy.

\section{Conclusion}
\setcounter{equation}{0}

In this paper we have proposed an improved energy functional 
for variational calculations in gauge field 
theories. This functional contains a non local term which approximately 
corresponds to the energy gain when projecting on gauge invariant states. 
This allows one to use  gaussian states as trial functionals and thus
perform analytic calculations for physical observables such as the 
chromoelectric and chromomagnetic susceptibilities, energy expectation values
and the gluon condensate.

The main purpose of this work was to check the ultraviolet behavior 
of our approximation scheme. By performing 
variational calculations near the perturbative 
vacuum we have shown that divergences can be eliminated 
by a renormalization of the coupling constant. This has
allowed us to recover the familiar one loop beta function  
in a way which makes transparent the various screening and 
antiscreening contributions.
In particular the screening term arises naturally
in our formalism, which was not the case in earlier
variational approaches \cite{BROWN1}.
We have also tested our variational method in the magnetic sector, checking 
that it reproduces the one-loop result by Savvidy \cite{SAVVIDY} for the
background field energy. This calculation provides us with  another 
derivation of the one-loop beta function.

Thus, our formalism appears to correctly reproduce the expected behavior
of non-Abelian theories in the ultraviolet sector. This strongly encourages
us to study its predictions in the non perturbative regime.
Indeed, as a variational approach, it is not at all 
restricted to the vicinity of the perturbative ground state, nor to small 
values of the coupling constant.
We would like to also emphasize that the Thouless-Valatin 
correction is the first 
step in an approximation scheme which can be constructed
systematically. Indeed, it is the first order term \cite{SCHUCK} in boson 
expansion methods which have been constructed by
Schwinger \cite{SCHWING2},  Dyson, \cite{DYSON}
Holstein-Primakoff \cite{HOLSTEIN},
Blaizot-Marshalek \cite{BLAIZOT-MAR}.

An attractive feature of the variational picture is that it allows one 
to treat situations where instabilities occur. This is the case when 
the trial state is not the lowest one and where saddle points are 
reached.
An example of such a situation is the occurence of negative 
modes generally found in the presence of a constant magnetic 
background field \cite{NIELSEN,OLESEN}. 
In this case it would be interesting to work 
out what is the optimal gaussian kernel in our approach. This state 
should be reached by allowing the variational space to include the subspace  
spanned by the negative modes.

Another attractive question is the investigation of the infrared 
behaviour. It  may provide some information on the confinement 
mechanism  and the generation of mass scales \cite{DIAKONOV}. 
Sum rules and the gluon condensate at finite
temperature also appear to be a promising field of investigation.

{\bf Acknowledgements}

  We are most grateful to A. K. Kerman for valuable remarks, 
regarding in particular the physical significance of the moment of inertia. 
Stimulating discussions with J. Polonyi and L. McLerran are also 
gratefully acknowledged.

\appendix

\section{Physical interpretation of section IV.B.
 in terms of the induced charge}

Let us show an other 
interpretation (and computation)
of the relation between $\rho^a$ and the external charge. What follows 
is directly inspired from 
 Gottfried and Weisskopf  in section IV.C of \cite{VIKI}.
Let us assume that the system is in presence of a given distribution of 
external charges 
$\rho_{ext}^a({\bf x})=\bfgrad\cdot{\bf E}_{ext}^a({\bf x})$ and
compute the corresponding  induced charge created by the quantum 
fluctuations of the gauge field ;
it is given by the mean value $<-f_{abc}A_i^b({\bf x})( E_L^i)^{c}({\bf x})>$ 
where
${\bf E}_L$ is the  longitudinal part of the 
chromoelectric field operator ${\bf E}$\cite{VIKI}.
This operator is a non-dynamical variable which is fixed by 
Gauss' law:
\begin{equation}
\bfgrad\cdot{\bf E}_L^a({\bf x})=
\rho_{ext}({\bf x})-f_{abc}A_i^b({\bf x})( E_L^i)^{c}({\bf x})
\end{equation}
which can be solved perturbatively, setting ${\bf E}^a_L({\bf x})=
{\bf E}^{a(0)}_L({\bf x})+
{\bf E}^{a(1)}_L({\bf x})
+{\bf E}^{a(2)}_L({\bf x})+...$. 
In fact, only the first two terms will be needed for a development 
of the total charge in first order 
of $g^2$. They verify the following set of equations :
\begin{eqnarray} \label{EINDUCED}
\bfgrad\cdot{\bf E}_L^{(0)a}({\bf x})&=& \rho^a_{ext}({\bf x})\nonumber\\
\bfgrad\cdot{\bf E}_L^{(1)a}({\bf x})&=& -f_{abc} A_i^b({\bf x})( E_L^i)^{(0)c}({\bf x}).
\end{eqnarray} 
The first equation shows that
${\bf E}_L^{(0)}({\bf x})={\bf E}_{ext}({\bf x})$. 
Then the second equation is solved by:
\begin{equation}
(E_L^i)^{(1)c}({\bf x})=  -\int \frac{{\rm d^3}y}{4 \pi }
\frac{x_i-y_i}{|{\bf x}-{\bf y}|^3}
f_{cde} A_j^d({\bf y}) (E_{ext}^j)^{e}({\bf y})
\end{equation}
The total charge reads therefore :
\begin{eqnarray} \label{TOTCHA}
\rho^a_{tot}({\bf x})&=&<{\cal G}^a({\bf x})>\nonumber\\
&=&\rho^a_{ext}({\bf x})+<-f_{abc} A_i^b({\bf x})
(E_{ext}^i)^{c}({\bf x})>\nonumber\\
&&\hskip 1cm +<-f_{abc} A_i^b({\bf x})( E_L^i)^{(1)c}({\bf x})>\nonumber\\
&\equiv&\rho^a_{ext}({\bf x})+\rho^a_{ind}({\bf x}).
\end{eqnarray}

The first term in $\rho^a_{ind}$
vanishes since linear in $A_i$, while the second term yields 
\begin{eqnarray} \label{RHO2}
 \rho_{ind}^a({\bf x})&\equiv& <-f_{abc} A_i^b({\bf x})
( E_L^i)^{(1)c}({\bf x})>_c\\
&= &f_{abc} f_{cde}
\int \frac{{\rm d^3}y}{4 \pi }
\frac{x_i-y_i}{|{\bf x}-{\bf y}|^3}\,
G^{bd}_{ij}({\bf x}-{\bf y})  (E_{ext})^e_j({\bf y}),\nonumber
\end{eqnarray}
or, after a Fourier transform,
\begin{equation} \label{8.4}
\rho_{ind}^a({\bf q})= f_{abc} f_{cde}
   (E_{ext})^e_j({\bf q})
\int \frac{{\rm d}^3x}{4 \pi }
e^{-i{\bf q}\cdot {\bf x}}\, 
\frac{x_i}{|{\bf x}|^3}\,
G^{bd}_{ij}({\bf x}).
\end{equation}
By also using
\begin{equation} \label{8.6}
\int \frac{{\rm d}^3x}{4 \pi }
e^{i{\bf k}.{\bf x}} 
\frac{x_i}{|{\bf x}|^3}=i\,\frac{k_i}{{\bf k}^2},
\end{equation}
we finally deduce:
\begin{equation} \label{8.7}
\rho_{ind}^a({\bf q})= f_{abc} f_{cde}
i   (E_{ext})^e_j({\bf q})
\int \frac{{\rm d}^3k}{(2 \pi)^3 }\,
\frac{(k-q)_i}{|{\bf k}+{\bf q}|^2}\,
G^{bd}_{ij}({\bf k}).
\end{equation}
Since $G$ is transverse this expression reduces to
\begin{equation} \label{8.8}
\rho_{ind}^a({\bf q})= -f_{abc} f_{cde}
i   (E_{ext})^e_j({\bf q})
\int \frac{{\rm d}^3k}{(2 \pi)^3 }
\frac{q_i}{|{\bf k}+{\bf q}|^2}
vG^{bd}_{ij}({\bf k}).
\end{equation}
Furthermore for a smooth charge distribution we can approximate  
${\bf k}+{\bf q}$ by  ${\bf k}$ in the above integral. Noting that the result
vanishes unless $i$=$j$ we have
\begin{equation} \label{8.9}
\rho_{ind}^a({\bf q})= 
\gamma  \rho_{ext}^a({\bf q}),
\end{equation}
with
\begin{equation} \label{8.10}
\gamma=g^2 C_N \frac{1}{6 \pi^2} 
\int_0^\infty \frac{{\rm d}k}{\varepsilon ({\bf k})}
= \delta.
\end{equation} 
The total charge is then given by 
\begin{eqnarray}
\rho^a_{tot}({\bf q})&=&\rho_{ext}^a({\bf q})+
\rho_{ind}^a({\bf q})\nonumber\\
&=&\rho_{ext}^a({\bf q})(1+\delta),
\end{eqnarray}
which is precisely  the  expression obtained in section IV.B
(cf. eq.~(\ref{RHO-RHOEXT})).

\section{ Proper-time calculation of the energy density}

Let us present here in some detail the calculation of the  quantity 
${\rm Tr}G^{-1}$ which enters the energy of the magnetic field in Sec. V.C. 
According to the Schwinger proper-time representation (\ref{REPSCHW}), one 
needs the matrix element $\langle {\bf x}|e^{-tK}|{\bf x}\rangle$.
Since, moreover, we are mainly interested in the ultraviolet behaviour of the 
energy, this expression is needed only at small values of $t$,
which allows us to perform expansions in powers of $t$ whenever necessary. 

As explained in Sec. V.C, we shall 
use the background field in eq.~(\ref{CHOIX}) for which:
\beq
 [{\cal D}_i,{\cal D}_j]^{ab}=-f^{ab3}\epsilon^{3ij}B
=T^3S_zB,\eeq
and therefore $[\Pi_i,\Pi_j]^{ab}=-[{\cal D}_i,
{\cal D}_j]^{ab}=-(T^3)_{ab}(S_z)_{ij}B$.
It is convenient to define the following operator (cf. eq.~(\ref{K2})):
\begin{equation} \label{TILDK}
{\tilde K}_{ij}\equiv K_{ij}+\Pi_i\Pi_j
\,=\,\Pi^2\delta_{ij}+2[\Pi_i,\Pi_j]\,,
\end{equation}
in terms of which $K_{ij}$ can be  rewritten as follows :
\begin{equation} 
K={\tilde K}{\hat Q}.
\end{equation}
By also using $[{\tilde K},{\hat Q}]=0$, and ${\hat Q}+{\hat P}=1$,
we deduce
\begin{equation} 
e^{-tK}=e^{-t{\tilde K}}{\hat Q}+{\hat P}.
\end{equation}
We thus have to compute the matrix element
$\langle {\bf x}|e^{-t{\tilde K}}{\hat Q}|{\bf x}\rangle$,
with : 
\begin{eqnarray}
{\tilde K}&=&\Pi^2-2T^{3}S_zB\nonumber\\
\langle {\bf x}|e^{-t{\tilde K}} |{\bf x}\rangle&=&\langle {\bf x}|e^{-t\Pi^2}
e^{2tT^3S_zB}|{\bf x}\rangle,
\end{eqnarray}
where the second line follows since $[\Pi^2,T^3S_z]=0$.

The computation of  $\langle {\bf x}|e^{-t\Pi^2}|{\bf x}\rangle$ is
well-known in the literature \cite{BREZIN}, with the result :
\begin{equation}\label{EQ1}
\langle {\bf x}|e^{-t\Pi^2}|{\bf x}\rangle=(\frac{1}{4\pi t})^{3/2}
\frac{tT^3B}{\sinh (tT^3B)}\end{equation}
We thus obtain:
\begin{equation}\label{EQ2}
\langle {\bf x}|e^{-t{\tilde K}} |{\bf x}\rangle=
(\frac{1}{4\pi t})^{3/2}
\frac{tT^3B}{\sinh (tT^3B)}e^{2tT^3S_zB}]\,.\end{equation}
By also using ${\hat Q}=1-{\hat P}$ and
$\Pi_i\Pi^2={\tilde K}\Pi_i$, one then rewrite
$\langle {\bf x}|e^{-t{\tilde K}}{\hat Q}|{\bf x}\rangle$ as :
\begin{equation}\label{BB1}
\langle {\bf x}|e^{-t{\tilde K}} |{\bf x}\rangle-
\langle {\bf x}|e^{-t{\tilde K}}\Pi_i\Pi_j
\frac{1}{{\tilde K}}|{\bf x}\rangle\end{equation}
The last term of this equation can be obtained from $\langle 
{\bf x}|e^{-t{\tilde K}}\Pi_i\Pi_j|{\bf x}\rangle$ by integration over t.

To calculate $\langle {\bf x}
|e^{-t{\tilde K}}\Pi_i\Pi_j|{\bf x}\rangle$, we follow Schwinger's method 
\cite{SCHWINGER} : We work in Heisenberg's
representation with $t=is$ and deduce
\begin{eqnarray}\label{PIJ}
\langle {\bf x}|e^{-t\Pi^2}\Pi_i\Pi_j
|{\bf x}\rangle&=&\langle {\bf x}|e^{-is\Pi^2}\Pi_i\Pi_j
|{\bf x}\rangle\nonumber\\
&=&\langle {\bf x}(s)|\Pi_i(0)\Pi_j(0)|{\bf x}(0)\rangle
\end{eqnarray}
where :
\begin{equation} \label{B2} \begin{array}{lll}
x_i(s)&=&e^{is\Pi^2}x_i(0)e^{-is\Pi^2} , \\
\Pi(s)&=&e^{is\Pi^2}\Pi(0)e^{-is\Pi^2}
\end{array}
\end{equation}
The operator $U(s)=e^{-is \Pi^2}$ can be interpreted as the evolution
operator of a particle governed by the Hamiltonian $\Pi^2$. We have 
\begin{eqnarray}
\frac{dx_i}{ds}&=&i[\Pi^2, x_i](s)\nonumber\\
&=&2\Pi(s) 
\end{eqnarray}
and
\begin{equation}
\frac{d\Pi}{ds}=i[\Pi^2,\Pi](s) 
\end{equation}
Using $
[\Pi^2,\Pi](s)=2[\Pi_k,\Pi_i]\Pi^k=-2i{\cal F}_{ik}\Pi^k$,
where
\begin{equation} 
{\cal F}_{ij}^{ab}\equiv i[{\cal D}_i,{\cal D}_j]^{ab}
=iT^3S_zB\end{equation}
one obtains
$\Pi(s)=(e^{2s{\cal F}})_{ik}\Pi^k(0)$, and thus :
\begin{eqnarray}
x_i(s)- x_i(0)&=&(\frac{e^{2s{\cal F}}-1}{{\cal F}})_{ij}\Pi^j(0)\nonumber\\
&\equiv&R_{ij}^{-1}\Pi^j(0)
\end{eqnarray}
The matrix element $\langle {\bf x}|e^{-t\Pi^2}\Pi_i\Pi_j
|{\bf x}\rangle$ can be now computed as (cf. eq.~(\ref{PIJ})):
\begin{eqnarray}
&&\langle {\bf x}(s)|\Pi_i(0)\Pi_j(0)|{\bf x}(0)\rangle\nonumber\\
&=&
\langle {\bf x}(s)|R_{im}R_{jn}( x_m(s)-x_m(0))(x_n(s)-x_n(0))|
{\bf x}(0)\rangle\nonumber\\
&=&R_{im}R_{jn}\langle {\bf x}(s)| x_m(s)x_n(s)- x_m(s)x_n(0)\nonumber\\
&&\hskip 1cm -x_n(s)
x_m(0)
+x_m(0) x_n(0)+[x_n(s),x_m(0)]|{\bf x}(0)\rangle\nonumber\\
&=&R_{im}R_{jn}R_{nk}^{-1}[\Pi_k(0),x_m(0)]\langle {\bf x}(s)|{\bf x}(0)\rangle
\nonumber\\
&=&-iR_{ij}\langle {\bf x}(s)|{\bf x}(0)\rangle
\end{eqnarray}                                                                   
Returning to the variable $s=-it$, wee obtain
\begin{equation} \langle {\bf x}|e^{-t\Pi^2}\Pi_i\Pi_j
|{\bf x}\rangle=(\frac{-i{\cal F}}{e^{-2it{\cal F}}-1})_{ij}\langle {\bf x}|e^{-t\Pi^2}
|{\bf x}\rangle,\end{equation}
and thus, finally,
\begin{eqnarray}
&&\langle {\bf x}|e^{-t{\tilde K}}\Pi_i\Pi_j|{\bf x}\rangle\nonumber\\
&=&\frac{T^3S_zB}
{2\sinh (tT^3S_zB)}(\frac{1}{4\pi t})^{3/2}
\frac{tT^3B}{\sinh (tT^3B)}e^{tT^3S_zB}\nonumber\\
&=&
\frac{1}{2t}(\frac{1}{4\pi
t})^{3/2}\Bigg\{1+T^3BS_zt\nonumber\\
&&\hskip 1.5cm +(\frac{(T^3)^2B^2S_z^2}{3}-
\frac{(T^3)^2B^2}{6})t^2+\ldots 
\Bigg\}
\end{eqnarray}
where in writing the second line
we have developed up to the second order in t.

The last equation can be now
integrated over $t$ to obtain the second term in (\ref{BB1}):
\begin{eqnarray}
\langle {\bf x}|e^{-t{\tilde K}}{\hat P}|{\bf x}\rangle
&=&
(\frac{1}{4\pi
t})^{3/2}\Bigg\{\frac{1}{3}+T^3BS_zt\\
&&+(\frac{(T^3)^2B^2}{6}-
\frac{(T^3)^2B^2S_z^2}{3})t^2
+\ldots \Bigg\}\nonumber
\end{eqnarray}
Developing $\langle {\bf x}|e^{-t{\tilde K}}|{\bf x}\rangle$ up to the second order
in t :
\begin{eqnarray}\label{SANSPRO}
\langle {\bf x}|e^{-t{\tilde K}}|{\bf x}\rangle&=&
(\frac{1}{4\pi
t})^{3/2}\Bigg\{1+2T^3BS_zt\\
&& +(2(T^3)^2B^2S_z^2-
\frac{(T^3)^2B^2}{6})t^2+\ldots \Bigg\},\nonumber
\end{eqnarray}
we deduce the following
expression for $U(t)\equiv\langle {\bf x}|e^{-t{\tilde K}}
{\hat Q}| {\bf x}
\rangle$ :
\begin{eqnarray}\label{UT}
U(t)&=&
(\frac{1}{4\pi
t})^{3/2}\Bigg\{\frac{2}{3}+T^3BS_zt\\
&&\hskip 0.5cm +(\frac{7}{3}(T^3)^2B^2S_z^2-
\frac{1}{3}(T^3)^2B^2)
t^2+\ldots \Bigg\}.\nonumber
\end{eqnarray}
Thus, finally,
\begin{eqnarray}\label{RES1}
{\rm Tr}\langle {\bf x}|G^{-1}| {\bf x}
\rangle&=&\frac{1}{\sqrt\pi}\int_{1/\Lambda^2}^{\infty}\frac{dt}{t^{3/2}}
(Tr\langle {\bf x}|{\hat Q}
|{\bf x}\rangle -TrU(t))\nonumber\\
&=&(\ldots
)\Lambda^4-\frac{C_N}{8\pi^2}(\frac{11}{3}B^2)
\ln\frac{\Lambda^2}{B}+{\cal O}(g^2),
\end{eqnarray}
where $\langle {\bf x}|{\hat Q}|{\bf x}\rangle $
has been obtained by setting $t=0$ in eq.~(\ref{UT}).
This is the result announced in eq.~(\ref{RES}).

\end{document}